\documentclass[conference, letterpaper]{IEEEtran}

\usepackage[letterpaper, top=1.92cm, bottom=3.0 cm, left=1.92cm, right=1.92cm]{geometry}
\IEEEoverridecommandlockouts
\usepackage{cite}
\usepackage{amsmath,amssymb,amsfonts}
\usepackage[numbers]{natbib}
\usepackage{textcomp}
\usepackage{xcolor}
\usepackage{amsmath}
\usepackage{caption}
\usepackage{soul}
\usepackage[most]{tcolorbox}
\usepackage{pgf}
\makeatletter
\def\thanks#1{\protected@xdef\@thanks{\@thanks
        \protect\footnotetext{#1}}}
\makeatother
\newcommand{\emphblockoption}{drop shadow,
  colframe=black!60,
  colback=black!10,
  coltitle=white!, 
    left=-15pt,
    right=15pt,
    boxrule=0pt,
    arc=-0pt}
\tcbset{emphblock/.style={code={\pgfkeysalsofrom{\emphblockoption}}}}

\usepackage[utf8]{inputenc} 
\usepackage{wrapfig}
\usepackage{url}            
\usepackage{booktabs}       
\usepackage{graphicx, subfigure}

\usepackage{pifont}

\newcommand{\leqnomode}{\tagsleft@true}
\newcommand{\reqnomode}{\tagsleft@false}

\usepackage{amsmath}

\usepackage{times}
\usepackage{url}

\usepackage{cite}
\usepackage{amsmath,amssymb,amsfonts}
\usepackage{graphicx}
\usepackage{textcomp}

\usepackage{dsfont,empheq}
\usepackage{mathrsfs}
\usepackage{algorithm}
\usepackage{algpseudocode}
\usepackage{algorithmicx}
\usepackage[textsize=tiny]{todonotes}
\usepackage{mathtools}
\usepackage{syntonly,bm,wrapfig}
\usepackage{cases,balance}

\usepackage{cleveref}

\usepackage{makecell}

\usepackage{amsthm,thmtools, thm-restate}

\newtheorem*{assumption*}{Assumption}

\usepackage{float}
\usepackage{placeins}
\usepackage{multirow}

\renewcommand\footnotemark{}

\graphicspath{{./figures/}}

\usepackage{amsmath,graphicx}

\def\BibTeX{{\rm B\kern-.05em{\sc i\kern-.025em b}\kern-.08em
    T\kern-.1667em\lower.7ex\hbox{E}\kern-.125emX}}
               
\begin{document}

\title{Leveraging Large Language Models for Wireless Symbol Detection via In-Context Learning}

\name{
{Momin Abbas \qquad Koushik Kar \qquad Tianyi Chen} 
\thanks{
The work of M. Abbas and T. Chen was supported by NSF CAREER 2047177, NSF ECCS 2412486, Cisco Research Award, and the IBM through the IBM-Rensselaer Future of Computing Research Collaboration.}
}
\address{Department of Electrical, Computer and Systems Engineering \\Rensselaer Polytechnic Institute, 
Troy, NY, USA}

\vspace{-0.2cm}
\maketitle

\begin{abstract}
Deep neural networks (DNNs) have made significant strides in tackling challenging tasks in wireless systems, especially when an accurate wireless model is not available. However, when available data is limited, traditional DNNs often yield subpar results due to underfitting. 
At the same time, large language models (LLMs) exemplified by GPT-3, have remarkably showcased their capabilities across a broad range of natural language processing tasks. 
\begin{center}
   \textsf{But how LLMs can benefit challenging {non-language tasks} in wireless systems is not fully unexplored.} 
\end{center}
In this work, we propose to leverage the \emph{in-context learning} ability (a.k.a. prompting) of LLMs to solve wireless tasks in the low data regime without any training or fine-tuning, unlike DNNs which require training. We further demonstrate that the performance of LLMs varies significantly when employed with different prompt templates. To solve this issue, we employ the latest LLM calibration methods. Our results reveal that using LLMs via ICL methods generally outperforms traditional DNNs on the symbol demodulation task and yields highly confident predictions when coupled with calibration techniques.
\end{abstract}

\vspace{-0.2cm}
\section{Introduction}
\label{sec:intro}
\subsection{Context and Motivation}
As the era of AI unfolds, it is expected that deep learning models will play a central role in shaping the future of wireless systems \cite{erpek2020deep}. 
Most work on AI in wireless communication leverages deep neural networks (DNNs) \cite{simeone2018very, dai2020deep, eldar2022machine, zhou2020deep}. 
To successfully integrate deep learning models into wireless systems, a key requirement is the ability to rapidly adapt to changing environmental conditions, even with limited information about the wireless systems \cite{simeone2020learning, chen2023learning}. This includes their ability to handle constantly changing wireless channel conditions using only a few pilot signals \cite{raviv2023modular}. 

DNN-based nonlinear channel predictors have been proposed through training of recurrent neural networks \cite{liu2006recurrent}, convolutional neural networks \cite{yuan2020machine}, and multi-layer perceptrons \cite{kim2020massive}. However, several studies, including \cite{jiang2020long, kim2020massive}, have reported that deep learning based predictors tend to require a large number of training data, while failing to outperform well-designed linear filters in the low-data regime.
This challenge becomes pronounced as neural networks increase in depth; see Table \ref{table:consolidated-performance}. This is critical in resource-constrained wireless systems, where the acquisition of data is expensive, necessitating costly hardware and skilled labor.

At the same time, despite significant advancements of Large Language Models (LLMs) in Natural Language Processing (NLP) and Computer Vision (CV) \citep{dong2019unified, GPT3}, pre-trained LLMs have faced limitations in their development within non-linguistic tasks, let alone wireless tasks. Therefore, combining wireless communications and natural language remains a challenge to utilize these capabilities.

In this work, we aim to achieve the best of both worlds by leveraging \emph{in-context learning} abilities of LLMs on the symbol detection task. We summarize contributions below:
\begin{itemize}
\itemsep-0.1em
  \item[\bf C1)] {We highlight the challenge in training traditional DNNs for symbol demodulation with limited data. To overcome this, we propose harnessing the in-context learning (ICL) abilities of LLMs through inference alone, without requiring any subsequent training or fine-tuning.}
 
  \item[\bf C2)] {As LLMs via ICL for wireless data is sensitive to changes in prompt templates, we propose employing state-of-the-art (SOTA) calibration methods \cite{zhao2021calibrate, abbas2024enhancing} designed for LLMs.}

  \item[\bf C3)] We empirically show that ICL methods generally outperform traditional DNNs in scenarios with limited data (e.g. 22\% performance improvement for 32-shots).
\end{itemize}

\vspace{-0.1cm}
\subsection{Related Work}
The majority of research in AI for communications relies on traditional frequentist learning methods that use traditional DNNs \cite{simeone2018very, dai2020deep, eldar2022machine}. These methods involve minimizing the (regularized) training loss, which serves as an estimate of the ground-truth population loss. However, in scenarios with limited data, this estimate becomes unreliable. Consequently, focusing on a single, optimized model parameter vector often results inaccurate and poorly calibrated probabilistic predictors, leading to overconfident decisions \cite{guo2017calibration, 9947031}.

Some methods focus on enhancing the calibration of DNNs through a validation-based post-processing phase. Platt scaling and temperature scaling \cite{platt1999, guo2017calibration} determine a fixed parametric mapping of the trained model output that minimizes the validation loss. In contrast, isotonic regression \cite{zadrozny2002transforming} utilizes a non-parametric binning approach. However, since these models primarily target either simple machine learning models or traditional DNNs, they often perform poorly in scenarios with limited data. \cite{cohen2023calibrating, angelopoulos2023conformal} examine how conformal prediction can be utilized as a general framework to ensure that AI models provide decisions with formal calibration guarantees. However, their notion of calibration differs significantly from ours. They transform probabilistic predictors into set predictors, where the set predictor is considered well-calibrated if it contains the correct output, and their goal does not prioritize performance improvement.

\begin{figure}[t]
\vspace{-0.2cm}
\begin{center}
    \includegraphics[width=0.45\textwidth]{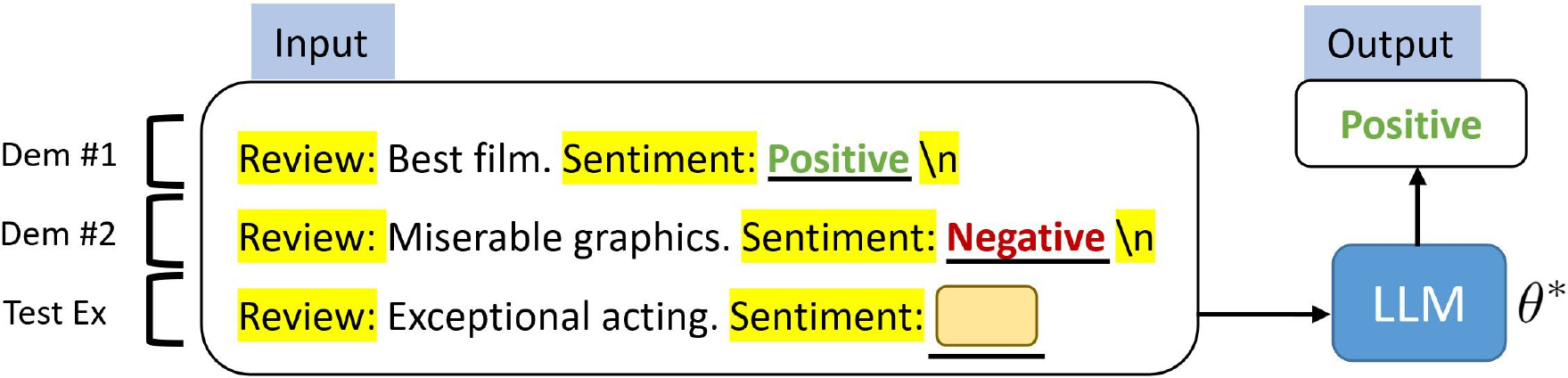}
\end{center}
\vspace{-0.1cm}
\caption{Example of ICL with LLM $\theta^*$.}
\label{ICL_fig}
\vspace{-0.4cm}
\end{figure}
Recently, LLMs such as GPT-3 \citep{GPT3} have showcased the capability of in-context learning. This feature enables a model to generate suitable outputs for a given query input by leveraging a prompt containing input-output example pairs tailored to the task at hand. ICL has proven to be highly effective in linguistic tasks with limited data, as it operates without the need for explicit training. However, ICL performance fluctuates across various prompt templates due to inadequate calibration \cite{zhao2021calibrate}. 
Recent studies have shown the potential of using transformer-type sequence models for MIMO detection tasks \cite{zecchin2024cell,zecchin2024context,rajagopalan2023transformers}. 
However, different from these works \cite{zecchin2024cell,zecchin2024context,rajagopalan2023transformers} that train a sequence model for wireless tasks and then employ ICL, {\em we employ ICL directly on the publically available LLMs in its organic form} and use advanced LLM calibration methods, as proposed in \citep{abbas2024enhancing, zhao2021calibrate}. This achieves high performance on the symbol detection problem while ensuring precise calibration of the LLMs.

\section{Formulation and Solution Approach} \label{problem_definition}
\vspace{-0.05cm}
In this section, we introduce the data model and then explore the difference between DNNs and LLMs.

\vspace{-0.05cm}
\subsection{Wireless Symbol Demodulation} \label{sym_demod_prob}
\vspace{-0.05cm}
We consider the wireless symbol demodulation problem from a discrete constellation, relying on received baseband signals that are susceptible to hardware imperfections, noise, and fading  \cite{7869303, 9947031, park2020learning}. Define $y_{i}$ as the $i$-th transmitted symbol, and $x_{i}$ as the corresponding received signal. Each transmitted symbol $y_{i}$ is drawn uniformly at random from a given constellation $Y$. We model $I/Q$ imbalance at the transmitter and phase fading as in \cite{cohen2023calibrating}. Accordingly, the ground-truth channel law connecting symbols $y_{i}$ into received samples $x_{i}$ is described by the equality
\begin{align} \label{eq7}
    x_{i} =  e^{j\Psi} f_{\rm IQ} (y_{i}) + v_{i},
\end{align}
for some random phase $\Psi \sim U[0,2\pi)$, $v_{i} \sim \mathcal{CN} (0, {\rm SNR}^{-1})$ and I/Q imbalance function $f_{\rm IQ}(y_{i}) = \Bar{y}^{\rm I}_{i} + j \Bar{y}^{\rm Q}_{i}$ \cite{4355276} where
\begin{align} \label{eq2}
   \begin{bmatrix}
        \Bar{y}^{\rm I}_{i} \\
        \Bar{y}^{\rm Q}_{i} 
   \end{bmatrix}
   =  
   \begin{bmatrix}
        1+ \epsilon & 0 \\
        0 & 1- \epsilon 
   \end{bmatrix}
   \begin{bmatrix}
        cos \delta & - \sin \delta \\
        - \sin \delta & cos \delta
   \end{bmatrix}
   \begin{bmatrix}
        {y}^{\rm I}_{i} \\
        {y}^{\rm Q}_{i} 
   \end{bmatrix}
\end{align}
where ${y}^{\rm I}_{i}$ and ${y}^{\rm Q}_{i}$ denote the real and imaginary parts of the modulated symbol $y_{i}$, and $\Bar{y}^{\mathrm{I}}_{i}$ and $\Bar{y}^{\mathrm{Q}}_{i}$ represent the real and imaginary parts of the transmitted symbol $f_{\mathrm{IQ}}(y_{i})$. In \eqref{eq2}, the channel state is defined by the tuple $(\Psi, \epsilon, \delta)$, encompassing the complex phase $\Psi$ and the $I/Q$ imbalance  $(\epsilon, \delta)$.

\subsection{Deep Neural Networks (DNNs)}
We consider a supervised learning setup with a dataset $\mathcal{D} = \{x_{i}, y_{i}\}_{i=1}^{N}$, consisting of $N$ examples represented as inputs $x_{i}$ corresponding outputs $y_{i}$. The goal is to make predictions for new, unseen test inputs $x_{\rm test}$ with an unknown output $y_{\rm test}$.

We are given a probabilistic predictor that implements a parametric conditional distribution model $ p(y_{\rm test} | x_{\rm test}, \phi)$ on the output $y_{\rm test} \in \mathcal{Y}$ from input $x_{\rm test} \in \mathcal{X}$, where $\phi \in \Phi$ denotes parameters of a DNN model. Given the training data set $\mathcal{D}$, the training algorithm produces an optimized $\phi_{\mathcal{D}}^{*}$.
For example, for a classification problem with $K$ labels (i.e. $|\mathcal{Y}|=K$), $p(y_{\rm test} | x_{\rm test}, \phi_{\mathcal{D}}^{*}) \in \mathbb{R}^{K}$ represents the last layer post-softmax probability vector.
We can then obtain a point prediction $\hat{y}_{\rm test}$ for output $y_{\rm test}$
given input $x_{\rm test}$ as the probability-maximizing output as
\begin{align} \label{eq_prediction}
    \hat{y}_{\rm test}(x_{\rm test} | \mathcal{D}) = \arg \max_{y_{\rm test}^{\prime} \in \mathcal{Y}} p(y_{\rm test}^{\prime}| x_{\rm test}, \phi_{\mathcal{D}}^{*}).
\end{align} 
However, this represents the conventional approach that \emph{uses training data to train a neural network} and then uses the trained model to make predictions on new test instances.

\subsection{Proposed Approach: LLM-based ICL (LMIC)} \label{icl_methods}
The most common method to leverage capabilities of LLMs is to fine-tune the LLM for specific tasks. However, fine-tuning LLMs can be problematic due to instability \cite{mosbach2020stability} caused by various hyperparameter configurations, leading to failed runs, unstable outcomes, and overfitting \citep{kumar2022fine}. Moreover, fine-tuning such large models can be costly and requires access to extensive data and the architecture and weights of LLMs, which may not be publicly available \citep{zhang2022opt}. 

Moreover, applying LLMs to non-language wireless tasks may require architecture adjustments, such as modifying input/output layers and loss functions \cite{dinh2022lift}. Therefore, it is natural to ask: \emph{Can we use LLMs for wireless tasks without altering the architecture or loss function}? We explore this question using the \emph{in-context learning} abilities of LLMs to solve wireless tasks. This approach offers a streamlined ``no-code machine learning" framework, enabling individuals with limited programming or machine learning expertise to address wireless tasks effortlessly.

To reduce lengthy fine-tuning processes and eliminate the need for accessing or modifying the model, recent advancements in LLMs, such as GPT-3, have showcased the capability of \emph{in-context learning}. ICL is a \emph{training-free} approach enabling the model to generate appropriate outputs for test samples by using prompts containing task-specific input-output examples. This approach works through an API without requiring direct access to the LLM. A visual representation of ICL is provided in Fig. \ref{ICL_fig}.

Specifically, ICL aims to predict a test sample $x_{\rm test}$ by conditioning on a prompt sequence $(f_{x}(x_{1}), f_{y}(y_{1}), \dots$, $f_{x}(x_{N}), f_{y}(y_{N}), f_{x}(x_{\rm test}))$. 
This sequence includes $N$-shot samples $\mathcal{D} = \{x_{i}, y_{i}\}_{i=1}^{N}$ (aka demonstrations) and the query test sample $x_{\rm test}$.
Here, $f_{x}(.)$ and $f_{y}(.)$ are template functions that provide predefined text descriptions for input and output, respectively (refer to text highlighted in \hl{yellow} in Fig. \ref{ICL_fig}).
Additionally, the output template function $f_{y}(.)$ may convert labels $y_{i}$ into natural language format instead of numeric/one-hot labels. For instance, in binary classification, it could transform labels $(0, 1)$ into (\textcolor{teal}{Positive}, \textcolor{purple}{Negative}) (see labels in Fig. \ref{ICL_fig}). Together, $f_{x}(.)$ and $f_{y}(.)$ constitute the \emph{prompt template}, providing a textual interpretation of the data. A prompt $P$ for an input $x_{\rm test}$ is defined as:
\begin{align} 
 \!\!\!    P(x_{\rm test}, (x_{i}, y_{i})_{i=1}^{N})  \triangleq d_{1} \oplus d_{2} \oplus \dots \oplus d_{N} \oplus f_x(x_{\rm test}) 
\end{align}
where each demonstration $d_i$ is given by $f_{x}(x_{i}) \oplus f_{y} (y_{i})$ and $\oplus$ denotes the concatenation operation.
\addtolength{\topmargin}{0.06 in}

Then for a pretrained LLM $M_{\phi^{*}}$ parameterized by $\phi^{*}$,
\begin{align}
    p_{\rm LLM}(y_{\rm test} | x_{\rm test}, \phi^{*}, \mathcal{D}) \triangleq M_{\phi^{*}}(P(x_{\rm test}, (x_{i}, y_{i})_{i=1}^{N})).
\end{align}
Note that the pre-trained LLM $M_{\phi^{*}}$ is \emph{not trained or fine-tuned} on the training data $\mathcal{D}$ and is therefore independent of $\mathcal{D}$. Instead, the training data serves as a `context' within the prompt, comprising a sequence of input-label pairs known as \emph{demonstrations}. Such a capability of an LLM to learn ``in-context'' presents an intriguing aspect whereby the LLM is capable of acquiring knowledge and performs well on a wide range of downstream tasks without any task-specific fine-tuning \cite{GPT3}. Moreover, for our approach, $x_{i}$ ($y_{i}$) represent the received (transmitted, respectively) signals (see Sec. \ref{sym_demod_prob}) and the corresponding prompt is shown in Table \ref{table_prompts}, where the bold numbers indicate the raw data for $x_{i}$ ($y_{i}$).

However, the dimension of $p_{\rm LLM}(y_{\rm test} | x_{\rm test}, \phi^{*}, \mathcal{D})$ is not $K$ as in traditional neural networks; instead, it corresponds to the number of tokens present in the vocabulary of the LLM; this is because LLMs perform next-token prediction, hence the dimension matches the number of tokens in LLM vocabulary.
Hence, we proceed by sampling tokens corresponding to the classes in the label space, resulting in a probability vector of size $K$. Finally, we can get the prediction following the same rule as Eq. \eqref{eq_prediction}, replacing $p(y_{\rm test}^{\prime} | x_{\rm test}, \phi^{*}_{\mathcal{D}})$ by $p_{\rm LLM}(y_{\rm test}^{\prime} | x_{\rm test}, \phi^{*}, \mathcal{D})$. We use this as our base method and refer to it as \emph{vanilla ICL}.

However, recent studies show that vanilla ICL's performance varies widely across different prompt templates and demonstrations \citep{abbas2024enhancing, zhao2021calibrate}, ranging from random guessing to state-of-the-art levels.
Additionally, we find that while these GPT-like models perform adequately via vanilla ICL, their predictions lack reliability on wireless tasks when assessed with Shannon entropy (see Figs. \ref{entropy_4/8}-\ref{entropy_16/32}). This observation aligns with the reliability concerns of LLMs identified in linguistic tasks \cite{abbas2024enhancing}. We gauge this is related to the poor calibration of LLMs \cite{zhao2021calibrate}. To address these reliability challenges posed by vanilla ICL, we leverage latest state-of-the-art (SOTA) calibration methods for LLMs, namely Contextual Calibration (ConC) \cite{zhao2021calibrate} and Linear Probe Calibration (LinC) \cite{abbas2024enhancing}.
Accuracy and calibration are independent criteria, with the presence of one not implying the other \cite{cohen2023calibrating}.

For brevity of notation, we denote the output probabilities $p_{\rm LLM}(y_{\rm test} | x_{\rm test}, \phi^{*}, \mathcal{D})$ as $\mathbf{p}$. Then, the goal is to linearly adjust the output probabilities using an affine transformation, also known as Platt Scaling \cite{platt1999}:
\begin{equation} \label{eq3}
     \Tilde{\mathbf{p}} = {\sf softmax} (\mathbf{A} \mathbf{p}+\mathbf{b}),
\end{equation}
where $\mathbf{A} \in \mathbb{R}^{K \times K}$ and $\mathbf{b} \in \mathbb{R}^{K}$ represent parameters applied to the original probabilities $\mathbf{p}$ to obtain new probabilities $\Tilde{\mathbf{p}}$. ConC then uses a prompt $P_{\rm cf} = P( \text{``N/A"}, (x_{i}, y_{i})_{i=1}^{N})$ (where test point $x_{\rm test}$ is replaced with a \emph{content-free} (cf) input such as the string “N/A” and obtain $\mathbf{p}$ for this content-free input, denoted by $\mathbf{p}_{\sf cf}$ i.e. $\mathbf{p}_{\sf cf} =  M_{\phi^{*}}(P_{\rm cf})$. The parameters are set via (i.e. no training needed)
\begin{align} 
     \mathbf{A} = {\sf diag}(\mathbf{p}_{\sf cf})^{-1} ~~~{\rm and}~~~\mathbf{b}= \mathbf{0}.
     \label{Eq_cf}
\end{align}
In contrast, LinC begins with a predefined set of calibration parameters, including zero initialization. Following this, LinC uses a few additional prompts to \emph{train} the matrix $\mathbf{A}$ and vector $\mathbf{b}$ before applying the affine transformation (for details, see \cite{abbas2024enhancing}). However, in our case, we do not use any additional samples and reuse the same $N$-shot demonstrations for training low-dimensional parameters $\mathbf{A}$ and $\mathbf{b}$.

\begin{table}
\centering
\scriptsize
\setlength{\tabcolsep}{2pt}
\begin{tabular}{cp{1.8cm}p{1.5cm}}
\toprule
\textbf{Prompt Template} &  \textbf{Label Space} \\
\midrule
8APSK signals are as follows:  & 0,\dots,7 \\
Signal 1's real part is \textbf{-2} and imaginary part is \textbf{4}. Actual Signal: \textbf{5} & (since $|Y|$=8) \\
Test Signal's real part is \textbf{3} and imaginary part is \textbf{-1}. Actual Signal: \\
\bottomrule
\end{tabular}
\vspace{-0.05cm}
\caption{The prompts template used for ICL methods; For brevity, here we show only one demonstration.}
\label{table_prompts}
\aboverulesep = 0.605mm
\belowrulesep = 0.984mm
\vspace{-0.5cm}
\end{table}


\begin{table}
\centering
\scriptsize
\setlength{\tabcolsep}{2pt}
\begin{tabular}{cp{7.5cm}p{2.5cm}}
\toprule
\textbf{Format\#} & \textbf{Prompt Template} \\
\midrule
\multirow{3}{*}{1} & 8APSK signals are as follows:   \\
& Signal 1's real part is -2 and imaginary part is 4. Actual Signal: 5  \\
& Test Signal's real part is 3 and imaginary part is -1. Actual Signal: \\
\midrule

\multirow{3}{*}{2} & 8APSK signals are as follows:   \\
& Signal 1's real part is -2 and imaginary part is 4. Actual Constellation: 5  \\
& Test Signal's real part is 3 and imaginary part is -1. Actual Constellation: \\
\midrule

\multirow{4}{*}{3} & 8APSK signals are as follows. Classify the signals based on the true set of classes [0, 1, 2, 3, 4, 5, 6, 7].   \\
& Signal 1's real part is -2 and imaginary part is 4. Actual Signal: 5  \\
& Test Signal's real part is 3 and imaginary part is -1. Actual Signal: \\
\midrule

\multirow{4}{*}{4} & Based on the 8APSK signals shown below, predict the Test Signal's output class from the set of classes [0, 1, 2, 3, 4, 5, 6, 7]:   \\
& Signal 1's real part is -2 and imaginary part is 4. Actual Signal: 5  \\
& Test Signal's real part is 3 and imaginary part is -1. Actual Signal: \\

\bottomrule
\end{tabular}
\caption{A list of different prompt templates that were used to investigate the impact of templates on Llama-2 7B 8-shot setting. For brevity, here we show only one demonstration.}
\label{table_prompts_ten}
\aboverulesep = 0.605mm
\belowrulesep = 0.984mm
\vspace{-0.5cm}
\end{table}

\section{Numerical Evaluation} \label{Sec_5}
We apply our LMIC approach, as described in Sec. \ref{icl_methods}, to address the symbol demodulation problem \cite{9250028, 7869303} in the presence of transmitter hardware imperfections. 
Unlike previous studies \cite{9947031, park2020learning}, which focused on frequentist and Bayesian learning via traditional DNNs, our goal is to use LMIC to achieve high accuracy and precise calibration under a severely limited resource regime (e.g., $< 50$ data samples), where traditional DNNs fail miserably (see Table \ref{table:consolidated-performance}).

Demodulation is implemented via an LLM $M_{\phi^{*}}$ as a next-token prediction problem (see Sec. \ref{icl_methods}). We used GPT-J \citep{wang2021gptj} with 6B parameters and two variants of the latest Llama-2 \citep{touvron2023llama} with 7B and 13B parameters\footnote{Note 13B model is the largest that can fit into our current GPU memory.}.

The last layer implements a softmax classification for the $K=|Y|$ possible constellation points. We employ the Amplitude-Phase-Shift-Keying (APSK) modulation with $K = 8$. The SNR level is set to SNR = 5 $\rm dB$. The amplitude and phase imbalance parameters are independent and distributed as $\epsilon \sim {\rm Beta} (\epsilon/0.15|5, 2)$ and $\delta \sim Beta(\delta/15^{\circ} | 5, 2)$, respectively \cite{park2020learning}. Unless specified otherwise, LMIC methods employ a fixed prompt template chosen manually to enhance performance, as demonstrated alongside examples in Table \ref{table_prompts}; bold numbers represent the raw data.

All our experiments were conducted on two NVIDIA RTX 3090 GPUs. As mentioned before, we consider the low resource regime, where the number of available training samples is scare (typically under $<$50). This is especially vital in resource-constrained scenarios where acquiring wireless data is expensive due to the costly hardware and skilled labor.

As baselines, we trained a fully connected deep neural network (DNN) similar to the one considered in \cite{9947031, cohen2023calibrating} with real inputs $x_{i}$ of dimension 2, following Eq. \eqref{eq7}. It consists of four hidden layers, with 10 neurons in the first hidden layer and 30 neurons in each subsequent hidden layer, each activated by ReLU. We also considered deeper networks with five, six and seven layers with 30 neurons in each additional hidden layer. The final layer performs softmax classification for the 
$|Y|$ possible constellation points.

To ensure a fair comparison, each DNN is trained using the identical set of samples employed as demonstrations within the prompt for LMIC methods. For instance, if there are 8-shots (i.e. demonstrations) in the prompt, the DNN baseline is trained using the same set of 8 samples.

Our main results are shown in Table \ref{table:consolidated-performance}. We observe that across most experiments (i.e., 15 out of 21 cases), our LMIC methods, particularly ConC and LinC, consistently demonstrate superior performance compared to the DNN baselines. This showcases the robust generalization capability of LLMs across various model sizes and few-shot settings. For instance, in the 32-shot experiment, Llama-2 7B outperforms the DNN-4 by a significant margin of about 22\% (69.31\% vs. 47.52\%).
Such a capability of ICL to understand contextual information from a handful of samples is particularly intriguing, especially when considering that the data is non-linguistic wireless data. We also note that the Llama-2 model outperforms the GPT-J model. This could be because Llama-2, released in August 2023, is one of the most recent models and therefore pre-trained on larger amounts of more recent data. 
We also notice that while the performance of DNNs generally declines with an increase in layers, eventually approaching near-guess accuracy with 7 hidden layers, the opposite trend is observed for LLMs: performance improves as the number of parameters increases (c.f. GPT-J 6B vs Llama-2 7/13B). Also, there is no consistent pattern in the performance of varying model sizes within the LLM family (i.e. Llama-2 7B vs 13B), which is consistent with previous works \cite{abbas2024enhancing}. However, Llama-2 7B notably achieves the highest accuracy of 69.31\% for 32-shots. Moreover, we observe that when the number of samples is less than the number of classes (i.e. $N < K$), DNNs with fewer hidden layers usually perform better than our LMIC methods (c.f. GPT-J and Llama-2 13, 5/6-shot results with DNN-4/5. This observation is in line with previous works that emphasize the pivotal role of label space in the success of ICL \cite{min2022rethinking}. 

\begin{table}[tb]
  \renewcommand{\arraystretch}{1.0}
  \normalsize
  \setlength\tabcolsep{2.9pt}
  \centering
  \begin{tabular}{ccllllllllll}
    \toprule
      Model & Vanilla ICL     & ConC   & LinC     \\
    \midrule
    
     Llama-2 7B &  0.2341 & 0.1166 & 0.1166 \\

    \bottomrule
  \end{tabular}
  
  \setlength{\abovecaptionskip}{0.2cm} 
  \caption{Expected Calibration Error (ECE) comparison between different ICL methods under 32-shot setting.}
  \label{table_ECE}
  \vspace{-0.5cm}
\end{table}

\begin{figure}[t]
\begin{center}
    \includegraphics[width=0.45\textwidth]{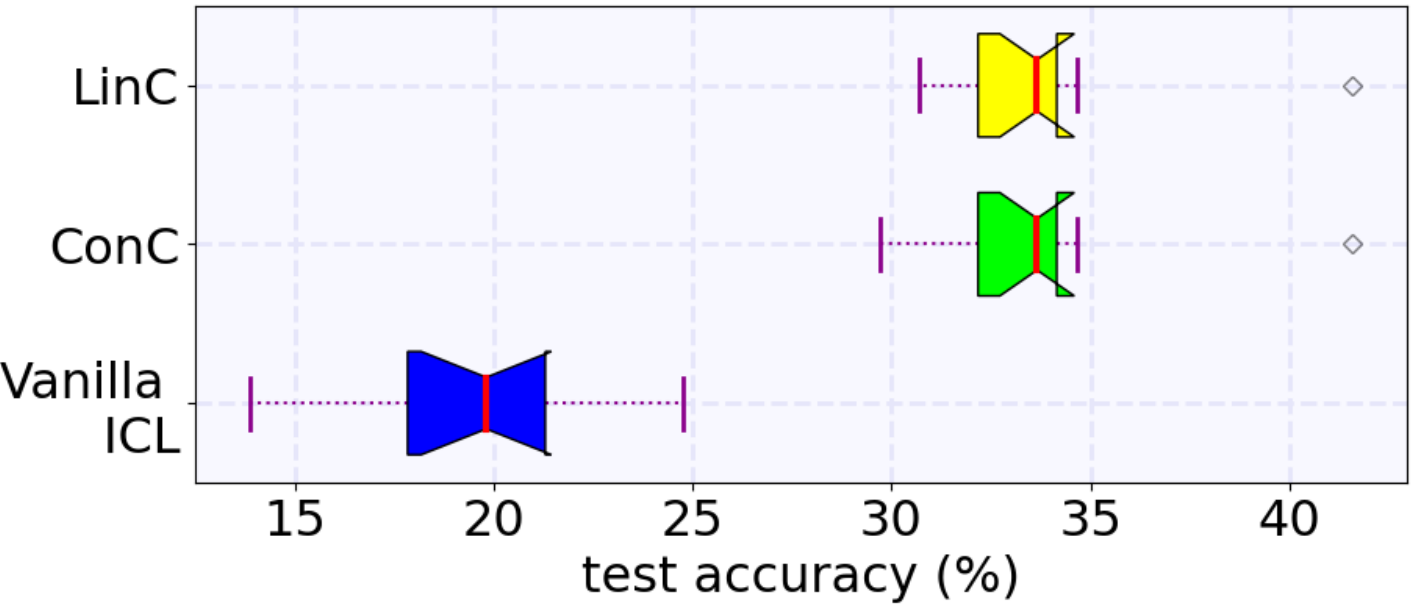}
\end{center}
\vspace{-0.3cm}
\caption{\small Comparison across ten different prompt templates.}
  \label{box_plot_template}
  \vspace{-0.2cm}
\end{figure}

As previously mentioned, prior works suggest that the performance of vanilla ICL fluctuates across different prompt templates in linguistic tasks. To investigate if this phenomenon also holds for non-linguistic wireless data, we use ten distinct prompt templates, four of which are listed in Table \ref{table_prompts_ten} (due to space constraints), on Llama-2 7B under 8-shot setting. From Fig. \ref{box_plot_template}, indeed the performance of vanilla ICL is volatile across different prompts while the latest SOTA calibration methods exhibit substantial improvements in accuracy with notably lower variance, highlighting the effectiveness of these methods in enhancing the model's performance across various prompt templates.

\begin{figure}[t]
\begin{center}
    \includegraphics[width=0.45\textwidth]{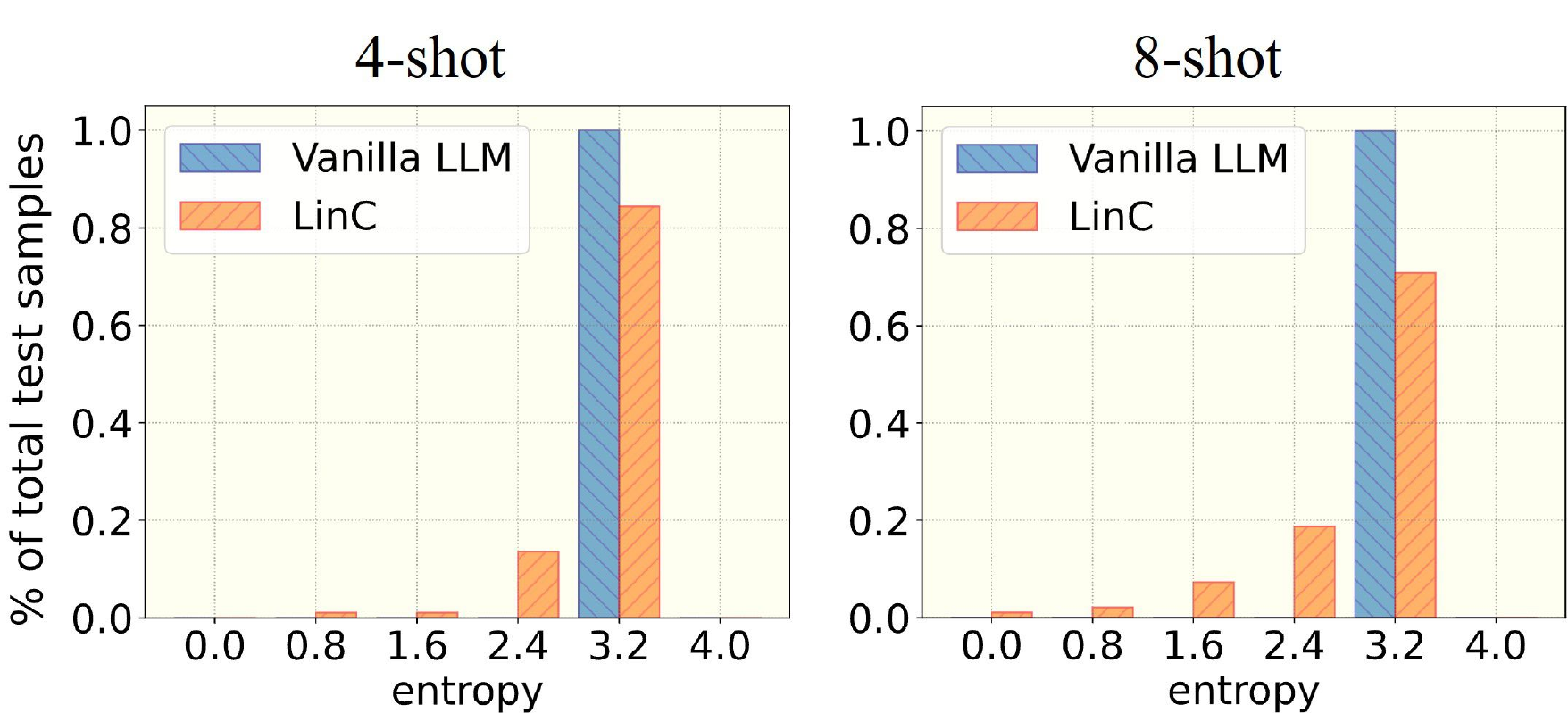}
\end{center}
\vspace{-0.3cm}
\caption{\small Shannon entropy histograms of different ICL methods on Llama-2 13B for 4/8-shots setting; we use logarithmic base two.}
  \label{entropy_4/8}
  \vspace{-0.3cm}
\end{figure}

\begin{figure}[t]
\vspace{-0.3cm}
\begin{center}
    \includegraphics[width=0.45\textwidth]{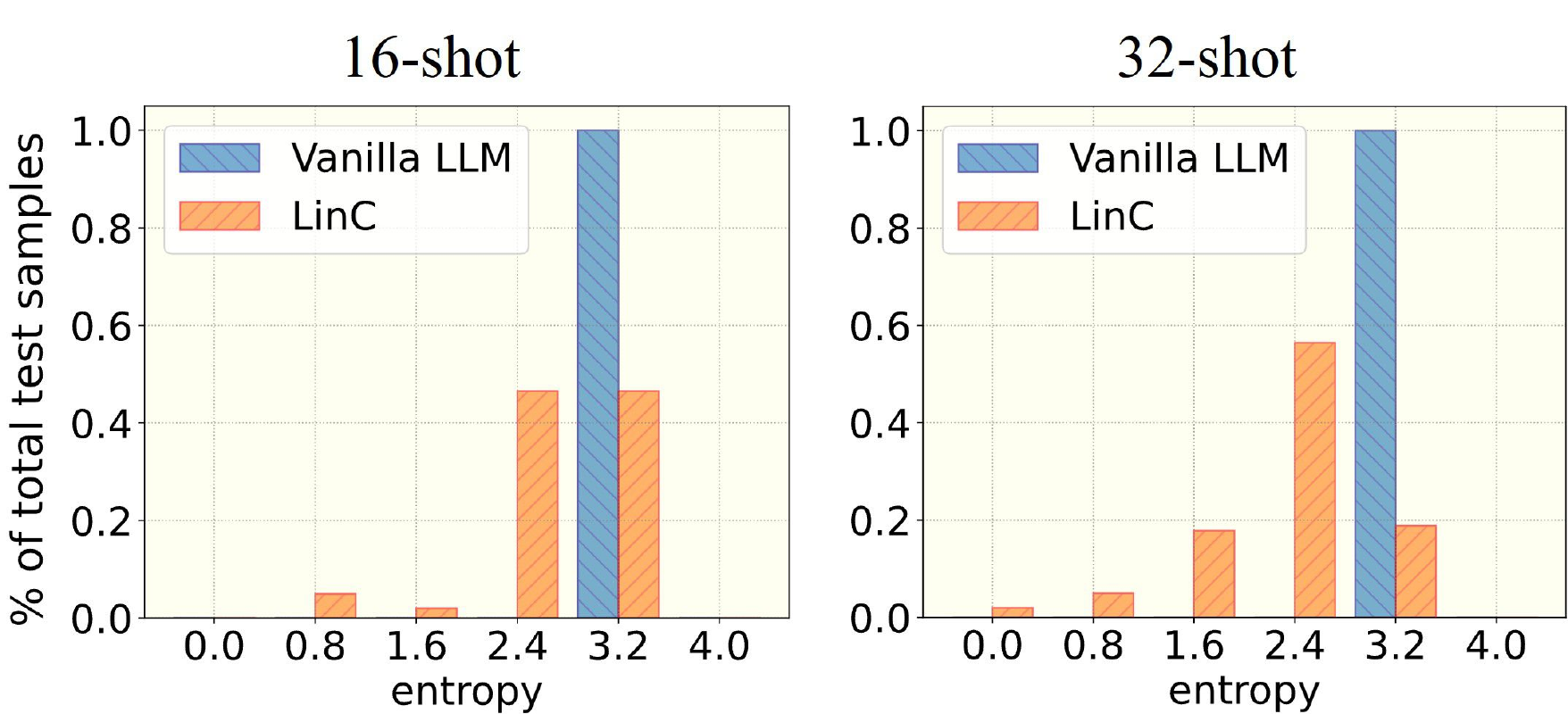}
\end{center}
\vspace{-0.3cm}
\caption{\small Shannon entropy histograms of different ICL methods on Llama-2 13B for 16/32-shot setting; we use logarithmic base two.}
  \label{entropy_16/32}
  \vspace{-0.1cm}
\end{figure}

\addtolength{\topmargin}{0.06 in}
We further evaluate the reliability of LLM predictions by employing the \textit{Shannon entropy} metric \citep{shannon1948mathematical}, which measures the expected uncertainty in a probability distribution $\mathbf{p}$).  A model is considered better when entropy values are lower. For this experiment we used our largest Llama-2 13B model to compare vanilla ICL and LinC, showing results for 4/8/16/32-shots in Figs. \ref{entropy_4/8}-\ref{entropy_16/32}. We oberve that employing vanilla ICL results in high entropy values, suggesting that most test predictions were made with very low confidence, indicating a tendency towards random guessing. These findings are consistent with prior studies for linguistic data \citep{abbas2024enhancing}.
In contrast, a calibrated LLM via LinC produces significantly lower entropy values, reflecting the increased confidence.

To further evaluate LLM calibration, we utilize the widely-used Expected Calibration Error (ECE) metric \citep{Naeini2015} to quantify the distance between predicted and actual probabilities. Table \ref{table_ECE} shows the results on Llama-2 7B, 32-shot setting (we only present this setting due to limited space). We observe that LLM calibration methods such as ConC and LinC consistently exhibit much lower ECE when compared to vanilla ICL, highlighting the critical importance of calibrating LLMs for wireless data.

Despite the merits of our approach, it has some limitations. Although our method eliminates the need for GPU-intensive training, it still relies on GPU memory for inference, albeit at a reduced scale compared to training or fine-tuning. Nevertheless, with the increasing adoption of LLMs across various domains, we anticipate that GPUs will become more readily accessible for deployment in wireless communication tasks in the future. Lastly, although we employed manually selected prompts and demonstrations, exploring how to integrate our framework with methods for selecting better demonstrations and prompt templates is an interesting future direction.

\begin{table}[!htp]
\centering
\footnotesize
\begin{tabular}{ccccccccc}
\cmidrule(r){1-8}
\textbf{Model} &  \multicolumn{7}{c}{\textbf{Shots}} \\
\cmidrule(r){1-8}
   &  4 & 5 & 6 & 8 & 16 & 24 & 32    \\
   \cmidrule(r){2-8}
   Guessing                                &  12.50  &  12.50  &  12.50  &  12.50  &  12.50  &  12.50  &  12.50           \\
   DNN-4                                  &  15.84  &  37.62  &  33.66  &  30.69  &  41.58  &  31.68  &  44.55           \\
   DNN-5                                  &  13.86  &  28.72  &  35.64  &  32.67  &  39.60  &  43.56  &  44.55           \\
   DNN-6                                  &  15.84  &  12.87  &  18.81  &  23.76  &  23.76  &  23.76  &  37.62           \\
   DNN-7                                  &  14.85  &  12.87  &  12.87  &  12.87  &  12.87  &  12.87  &  25.74           \\
   GPT-J 6B$^*$                           &  19.79  &  18.81  &  15.84  &  16.83  &  37.62  &  22.77  &  18.81           \\
   GPT-J 6B$^\dag$                        &  24.75  &  33.66  &  27.72  &  33.66  &  41.58  &  47.52  &  41.58           \\
   GPT-J 6B$^\ddag$                       &  24.75  &  33.66  &  28.71  &  35.64  &  43.56  &  46.53  &  41.58           \\ 

 \cmidrule(r){1-8}

    DNN-4                                 &  26.73  &  31.68  &  41.58  &  37.62  &  40.59  &  50.50  &  47.52           \\
    DNN-5                                 &  23.76  &  24.75  &  40.59  &  33.66  &  39.60  &  52.48  &  44.55           \\
    DNN-6                                 &  21.78  &  13.86  &  25.74  &  20.79  &  12.87  &  39.60  &  24.75           \\
    DNN-7                                 &  11.88  &  12.87  &  12.87  &  12.87  &  12.87  &  22.77  &  12.87           \\
    Llama-7B$^*$                          &  33.33  &  20.79  &  26.73  &  16.83  &  31.68  &  58.42  &  64.36           \\
    Llama-7B$^\dag$                        &  29.17  &  40.59  &  39.60  &  41.58  &  49.50  &  59.41  &  69.31           \\
    Llama-7B$^\ddag$                        &  29.17  &  40.59  &  39.60  &  41.58  &  49.50  &  58.42  &  69.31           \\ 

 \cmidrule(r){1-8}

    DNN-4                                 &  31.68  &  39.60  &  36.63  &  32.67  &  47.52  &  59.40  &  45.54          \\
    DNN-5                                 &  28.71  &  22.77  &  23.76  &  34.65  &  32.67  &  45.54  &  50.50          \\
    DNN-6                                 &  22.77  &  25.74  &  19.80  &  21.78  &  22.77  &  43.56  &  30.69          \\
    DNN-7                                 &  21.78  &  12.87  &  17.82  &  20.79  &  20.79  &  12.87  &  12.87          \\
    Llama-13B$^*$                         &  26.04  &  18.81  &  31.68  &  31.25  &  40.59  &  54.46  &  53.47          \\
    Llama-13B$^\dag$                       &  37.50  &  32.67  &  33.66  &  38.54  &  49.50  &  65.35  &  58.42          \\
    Llama-13B$^\ddag$                      &  37.50  &  32.67  &  33.66  &  38.54  &  49.50  &  65.35  &  58.42          \\ 
   
 \cmidrule(r){1-8}
\end{tabular}
\caption{Performance comparison for the system demodulation task; \{\} in DNN-\{\} refers to the number of hidden layers of the fully-connected deep neural network; $^*$ denotes vanilla ICL, $^\dag$ denotes ConC, and $^\ddag$ denotes LinC.}
\label{table:consolidated-performance}
\vspace{-0.6cm}
\end{table}

\section{Conclusions}
While LLMs have been extensively studied for linguistic tasks, their utilization for non-linguistic wireless data remains largely unexplored. In this work, we capitalize on the \emph{in-context learning} abilities of LLMs that not only achieves high performance but also yields highly confident predictions when integrated with SOTA LLM calibration techniques, especially in data-scarce scenarios where traditional DNNs typically fail. We believe these findings carry important implications for advancing wireless systems through large language models. 
In the future, our aim is to investigate the high-resource regime, utilizing abundant data and compute to first fine-tune LLMs and then employ our LMIC approach.

{\fontsize{8.5}{10}\selectfont 
\bibliographystyle{IEEEbib}
\bibliography{myabrv,bmaml,maml_theory,maml,other,bilevel,ensemble, refs}

\begin{thebibliography}{10}

\bibitem{erpek2020deep}
Tugba Erpek, Timothy~J O’Shea, Yalin~E Sagduyu, Yi~Shi, and T~Charles Clancy,
\newblock ``Deep learning for wireless communications,''
\newblock {\em Development and Analysis of Deep Learning Architectures}, pp. 223--266, 2020.

\bibitem{simeone2018very}
Osvaldo Simeone,
\newblock ``A very brief introduction to machine learning with applications to communication systems,''
\newblock {\em IEEE Trans. on Cognitive Comm. and Netw.}, vol. 4, no. 4, pp. 648--664, 2018.

\bibitem{dai2020deep}
Linglong Dai, Ruicheng Jiao, Fumiyuki Adachi, H~Vincent Poor, and Lajos Hanzo,
\newblock ``Deep learning for wireless communications: An emerging interdisciplinary paradigm,''
\newblock {\em IEEE Wireless Communications}, vol. 27, no. 4, pp. 133--139, 2020.

\bibitem{eldar2022machine}
Yonina~C Eldar, Andrea Goldsmith, Deniz G{\"u}nd{\"u}z, and H~Vincent Poor,
\newblock {\em Machine learning and wireless communications},
\newblock Cambridge University Press, 2022.

\bibitem{zhou2020deep}
Ruolin Zhou, Fugang Liu, and Christopher~W Gravelle,
\newblock ``Deep learning for modulation recognition: A survey with a demonstration,''
\newblock {\em IEEE Access}, vol. 8, pp. 67366--67376, 2020.

\bibitem{simeone2020learning}
Osvaldo Simeone, Sangwoo Park, and Joonhyuk Kang,
\newblock ``From learning to meta-learning: Reduced training overhead and complexity for communication systems,''
\newblock in {\em 2020 2nd 6G Wireless Summit (6G SUMMIT)}. IEEE, 2020, pp. 1--5.

\bibitem{chen2023learning}
Lisha Chen, Sharu~Theresa Jose, Ivana Nikoloska, Sangwoo Park, Tianyi Chen, Osvaldo Simeone, et~al.,
\newblock ``Learning with limited samples: Meta-learning and applications to communication systems,''
\newblock {\em Foundations and Trends{\textregistered} in Signal Processing}, vol. 17, no. 2, pp. 79--208, 2023.

\bibitem{raviv2023modular}
Tomer Raviv, Sangwoo Park, Osvaldo Simeone, and Nir Shlezinger,
\newblock ``Modular model-based bayesian learning for uncertainty-aware and reliable deep mimo receivers,''
\newblock in {\em IEEE ICC Workshops}, 2023.

\bibitem{liu2006recurrent}
Wei Liu, Lie-Liang Yang, and Lajos Hanzo,
\newblock ``Recurrent neural network based narrowband channel prediction,''
\newblock in {\em Proc. IEEE 63rd Vehicular Technology Conference}, Melbourne, Australia, 2006, vol.~5, pp. 2173--2177.

\bibitem{yuan2020machine}
Jide Yuan, Hien~Quoc Ngo, and Michail Matthaiou,
\newblock ``Machine learning-based channel prediction in massive mimo with channel aging,''
\newblock {\em IEEE Transactions on Wireless Communications}, vol. 19, no. 5, pp. 2960--2973, 2020.

\bibitem{kim2020massive}
Hwanjin Kim, Sucheol Kim, Hyeongtaek Lee, Chulhee Jang, Yongyun Choi, and Junil Choi,
\newblock ``Massive mimo channel prediction: Kalman filtering vs. machine learning,''
\newblock {\em IEEE Transactions on Communications}, vol. 69, no. 1, pp. 518--528, 2020.

\bibitem{jiang2020long}
Wei Jiang, Mathias Strufe, and Hans~Dieter Schotten,
\newblock ``Long-range mimo channel prediction using recurrent neural networks,''
\newblock in {\em Proc. IEEE Annual Consumer Communications \& Networking Conference}, Las Vegas, NV, 2020, pp. 1--6.

\bibitem{dong2019unified}
Li~Dong, Nan Yang, Wenhui Wang, Furu Wei, Xiaodong Liu, Yu~Wang, Jianfeng Gao, Ming Zhou, and Hsiao-Wuen Hon,
\newblock ``Unified language model pre-training for natural language understanding and generation,''
\newblock in {\em Advances in Neural Information Processing Systems}, 2019, vol.~32.

\bibitem{GPT3}
Tom Brown, Benjamin Mann, Nick Ryder, Melanie Subbiah, Jared~D Kaplan, Prafulla Dhariwal, Arvind Neelakantan, Pranav Shyam, Girish Sastry, Amanda Askell, et~al.,
\newblock ``Language models are few-shot learners,''
\newblock {\em Advances in neural information processing systems}, vol. 33, pp. 1877--1901, 2020.

\bibitem{zhao2021calibrate}
Zihao Zhao, Eric Wallace, Shi Feng, Dan Klein, and Sameer Singh,
\newblock ``Calibrate before use: Improving few-shot performance of language models,''
\newblock in {\em International Conference on Machine Learning}, 2021, pp. 12697--12706.

\bibitem{abbas2024enhancing}
Momin Abbas, Yi~Zhou, Parikshit Ram, Nathalie Baracaldo, Horst Samulowitz, Theodoros Salonidis, and Tianyi Chen,
\newblock ``Enhancing in-context learning via linear probe calibration,''
\newblock in {\em International Conference on Artificial Intelligence and Statistics}, 2024.

\bibitem{guo2017calibration}
Chuan Guo, Geoff Pleiss, Yu~Sun, and Kilian~Q Weinberger,
\newblock ``On calibration of modern neural networks,''
\newblock in {\em International conference on machine learning}, 2017, pp. 1321--1330.

\bibitem{9947031}
Kfir~M. Cohen, Sangwoo Park, Osvaldo Simeone, and Shlomo Shamai,
\newblock ``Bayesian active meta-learning for reliable and efficient ai-based demodulation,''
\newblock {\em IEEE Transactions on Signal Processing}, vol. 70, pp. 5366--5380, 2022.

\bibitem{platt1999}
John Platt,
\newblock ``Probabilistic outputs for support vector machines and comparisons to regularized likelihood methods,''
\newblock {\em Adv. Large Margin Classif.}, vol. 10, 06 2000.

\bibitem{zadrozny2002transforming}
Bianca Zadrozny and Charles Elkan,
\newblock ``Transforming classifier scores into accurate multiclass probability estimates,''
\newblock in {\em Proceedings of the eighth ACM SIGKDD international conference on Knowledge discovery and data mining}, 2002, pp. 694--699.

\bibitem{cohen2023calibrating}
Kfir~M Cohen, Sangwoo Park, Osvaldo Simeone, and Shlomo~Shamai Shitz,
\newblock ``Calibrating ai models for few-shot demodulation via conformal prediction,''
\newblock in {\em IEEE International Conference on Acoustics, Speech and Signal Processing}, 2023.

\bibitem{angelopoulos2023conformal}
Anastasios~N Angelopoulos, Stephen Bates, et~al.,
\newblock ``Conformal prediction: A gentle introduction,''
\newblock {\em Foundations and Trends{\textregistered} in Machine Learning}, vol. 16, no. 4, pp. 494--591, 2023.

\bibitem{zecchin2024cell}
Matteo Zecchin, Kai Zu, and Osvaldo Simeone,
\newblock ``Cell-free multi-user mimo equalization via in-context learning,''
\newblock {\em arXiv preprint:2404.05538}, 2024.

\bibitem{zecchin2024context}
Matteo Zecchin, Kai Yu, and Osvaldo Simeone,
\newblock ``In-context learning for mimo equalization using transformer-based sequence models,''
\newblock in {\em IEEE ICC Workshops}, 2024.

\bibitem{rajagopalan2023transformers}
Vicram Rajagopalan, Vishnu~Teja Kunde, Chandra Shekhara~Kaushik Valmeekam, Krishna Narayanan, Srinivas Shakkottai, Dileep Kalathil, and Jean-Francois Chamberland,
\newblock ``Transformers are efficient in-context estimators for wireless communication,''
\newblock {\em arXiv preprint:2311.00226}, 2023.

\bibitem{7869303}
Ahmed~G. Helmy, Marco Di~Renzo, and Naofal Al-Dhahir,
\newblock ``On the robustness of spatial modulation to i/q imbalance,''
\newblock {\em IEEE Communications Letters}, vol. 21, no. 7, pp. 1485--1488, July 2017.

\bibitem{park2020learning}
Sangwoo Park, Hyeryung Jang, Osvaldo Simeone, and Joonhyuk Kang,
\newblock ``Learning to demodulate from few pilots via offline and online meta-learning,''
\newblock {\em IEEE Transactions on Signal Processing}, vol. 69, pp. 226--239, 2020.

\bibitem{4355276}
Deepaknath Tandur and Marc Moonen,
\newblock ``Joint adaptive compensation of transmitter and receiver iq imbalance under carrier frequency offset in ofdm-based systems,''
\newblock {\em IEEE Transactions on Signal Processing}, vol. 55, no. 11, pp. 5246--5252, 2007.

\bibitem{mosbach2020stability}
Marius Mosbach, Maksym Andriushchenko, and Dietrich Klakow,
\newblock ``On the stability of fine-tuning {\{}bert{\}}: Misconceptions, explanations, and strong baselines,''
\newblock in {\em International Conference on Learning Representations}, 2021.

\bibitem{kumar2022fine}
Ananya Kumar, Aditi Raghunathan, Robbie~Matthew Jones, Tengyu Ma, and Percy Liang,
\newblock ``Fine-tuning can distort pretrained features and underperform out-of-distribution,''
\newblock in {\em International Conference on Learning Representations}, 2022.

\bibitem{zhang2022opt}
Susan Zhang, Stephen Roller, Naman Goyal, Mikel Artetxe, Moya Chen, Shuohui Chen, Christopher Dewan, Mona Diab, Xian Li, Xi~Victoria Lin, et~al.,
\newblock ``Opt: Open pre-trained transformer language models,''
\newblock {\em arXiv preprint:2205.01068}, 2022.

\bibitem{dinh2022lift}
Tuan Dinh, Yuchen Zeng, Ruisu Zhang, Ziqian Lin, Michael Gira, Shashank Rajput, Jy-yong Sohn, Dimitris Papailiopoulos, and Kangwook Lee,
\newblock ``Lift: Language-interfaced fine-tuning for non-language machine learning tasks,''
\newblock in {\em Advances in Neural Information Processing Systems}, 2022, pp. 11763--11784.

\bibitem{9250028}
Yi~Zhang, Akash Doshi, Rob Liston, Wai-Tian Tan, Xiaoqing Zhu, Jeffrey~G. Andrews, and Robert~W. Heath,
\newblock ``Deepwiphy: Deep learning-based receiver design and dataset for ieee 802.11ax systems,''
\newblock {\em IEEE Transactions on Wireless Communications}, 2021.

\bibitem{wang2021gptj}
Ben Wang and Aran Komatsuzaki,
\newblock ``Gpt-j-6b: A 6 billion parameter autoregressive language model,'' 2021.

\bibitem{touvron2023llama}
Hugo Touvron, Louis Martin, Kevin Stone, Peter Albert, Amjad Almahairi, Yasmine Babaei, Nikolay Bashlykov, Soumya Batra, Prajjwal Bhargava, and Shruti Bhosale,
\newblock ``Llama 2: Open foundation and fine-tuned chat models,''
\newblock {\em arXiv preprint:2307.09288}, 2023.

\bibitem{min2022rethinking}
Sewon Min, Xinxi Lyu, Ari Holtzman, Mikel Artetxe, Mike Lewis, Hannaneh Hajishirzi, and Luke Zettlemoyer,
\newblock ``Rethinking the role of demonstrations: What makes in-context learning work?,''
\newblock {\em arXiv preprint:2202.12837}, 2022.

\bibitem{shannon1948mathematical}
Claude~E Shannon,
\newblock ``A mathematical theory of communication,''
\newblock {\em The Bell system technical journal}, vol. 27, no. 3, pp. 379--423, 1948.

\bibitem{Naeini2015}
Mahdi Pakdaman~Naeini, Gregory Cooper, and Milos Hauskrecht,
\newblock ``Obtaining well calibrated probabilities using bayesian binning,''
\newblock in {\em Proc. of AAAI Conference on Artificial Intelligence}, 2015.

\end{thebibliography}
}

\end{document}